# Content-Based Video Browsing by Text Region Localization and Classification

Bassem Bouaziz, Walid Mahdi. Tarek Zlitni, Abdelmajid ben Hamadou,

*MIRACL, Multimedia InfoRmation system and Advanced Computing Laboratory Higher Institute of Computer Science and Multimedia, Sfax, BP 1030 - Tunisia*

*Abstract*—**The amount of digital video data is increasing over the world. It highlights the need for efficient algorithms that can index, retrieve and browse this data by content. This can be achieved by identifying semantic description captured automatically from video structure. Among these descriptions, text within video is considered as rich features that enable a good way for video indexing and browsing. Unlike most video text detection and extraction methods that treat video sequences as collections of still images, we propose in this paper spatiotemporal video-text localization and identification approach which proceeds in two main steps: text region localization and text region classification. In the first step we detect the significant appearance of the new objects in a frame by a split and merge processes applied on binarized edge frame pair differences. Detected objects are, a priori, considered as text. They are then filtered according to both local contrast variation and texture criteria in order to get the effective ones. The resulted text regions are classified based on a visual grammar descriptor containing a set of semantic text class regions characterized by visual features. A visual table of content is then generated based on extracted text regions occurring within video sequence enriched by a semantic identification. The experimentation performed on a variety of video sequences shows the efficiency of our approach.**

## I. INTRODUCTION

Content-based indexing, browsing and retrieval of digital video and media archives are challenging problems and are rapidly gaining widespread research and commercial interest. To enable users to quickly locate their interested

B.Bouaziz. Multimedia Information systems and Advanced Computing Laboratory, Sfax, Tunisia. Phone 00216 98 48 80 63; fax : 00216 74 86 23 32; e-mail :bassem.bouaziz@fsegs.rnu.tn; bassem.bouaziz@gmail.com.

W. Mahdi, Multimedia Information systems and Advanced Computing Laboratory, Sfax, Tunisia. Fax: 00216 74 86 23 32; e-mail: walid.mahdi@isimsf.rnu.tn.

T. Zlitni, Multimedia Information systems and Advanced Computing Laboratory, Sfax, Tunisia. Fax: 00216 74 86 23 32; e-mail: tarek.zlitni@isimsf.rnu.tn.

A. Ben Hamadou, Multimedia Information systems and Advanced Computing Laboratory, Sfax, Tunisia. Fax: 00216 74 86 23 32; e-mail: abdelmajid.bennhamadou@isimsf.rnu.tn.

content in an enormous volume of video data, the extraction of content descriptive features is required. However, it is well recognized that low-level features as measures of color, texture and shape [4] are not enough for uniquely discriminating across different video content. Extracting more descriptive features and higher level entities as text [24] or faces is of great value in designing systems for efficiently browsing and retrieving the video content. Among all these features, text is most reliable for this purpose especially the superimposed text. Since it is generated by video title machines, it can easily express abundant information highly related to the current video content. Indeed, in TV broadcasting superimposed text carries important information to the viewer. In news broadcast, the topic of news event or scene location and dates are sometimes displayed. In sports, game scores event and players' names are displayed from time to time on the screen. If these text occurrences could be detected, extracted and identified automatically, they would be a valuable source of high-level semantics for indexing and browsing [10, 39, 24, 7, 44, 43]. This text could be also used for MPEG-7 descriptor based application [9, 1]. Unlike Moving texts like rolling text news, which usually less indicates the video content, most of superimposed texts containing useful description information are static. Therefore our work focuses on static superimposed text. However, extracting superimposed text embedded in video frames is not a trivial task. The difficulty comes from the variation of font-size, font-color, spacing, contrast and mostly the background complexity. In addition, popular lossy compression methods such as MPEG often lower the video frame quality and consequently the quality of text stroke. Obviously, the superimposed text has several characteristics [17], in particular the contrast, the alignment, the directionality, the regularity and the temporal redundancy i.e. the same text region appears over several contiguous frames at the same positions. Although several text detection and localization approaches process on video frames, they usually treat each frame as an independent and still image. Consequently, the detection and localization will be performed for every frame of a video or by a predefined step (each four frame for example). Such approaches require a matching process to remove duplicate detected regions. Given the large number of video frames, the computational cost may be very important. As continuity of our previous works [6, 5], we propose in this paper a spatio-temporal approach for video text localization and classification considering the characteristics of text mentioned above. The main idea consists on locating the frame where text regions appear and extracting the text features from the frame difference. The already located text region are then extracted and classified according to a visual grammar descriptor. The rest of the paper is organized as follows. Section 2 reviews related works for the text detection and localization. In section 3 we detail our text region localization and extraction approach composed of two main steps: potential text localization and resulted text region filtering. In section 4, we describe the classification of automatically extracted text regions with the use of visual grammar descriptors. The experimental results are presented in section 5. Finally, concluding remarks and further work directions are depicted in the last section.

## II. RELATED WORK

In the field of content-based information retrieval, video text detection and extraction has drawn much attention of many researchers. Existing text detection methods can be classified into color based methods [25, 45, 3, 16, 2, 11, 27], texture classification methods [30, 19, 12, 15, 37, 20, 42, 21, 34, 47, 23] and edge detection methods. [38, 24, 26, 22, 7, 26, 40, 38, 43, 18, 46, 33]. The first category assumes that the text regions have uniform colors or intensity and satisfy certain constraints as size and shape. Before performing a connected component analysis, they perform in general a color reduction and segmentation in some selected color channel as the red channel in [2] or in color space as Lab space chosen as in [25]. Jain et al. [16] proposed a detection method to deal with color images. It calculates the similarity of different color values and uses color reduction method to enhance the speed of similarity calculation. The similar regions are merged after color quantization to fill out the text region. Besides, Pei et al. [27] proposed neural network color quantization for text detection. Yi et al [45] proposed a Homogeneous color based and sharp edges method. Color-based clustering is employed to decompose the color edge map of image into several edge maps. [11] Gatos et al proposed a method based on adaptative binarization of gray level image and the associated inverse. Before performing the connected component analyses, a decision function is called to determine which image contains the textual information. Guo et al [13] used a histogram based color clustering within a refinement step of text localization. The color space dimension is reduced to $16 \times 16 \times 16$ cell after corner detection according to pyramid decomposition of gray level image and morphologic operation to connect region. This method detect only horizontal text and sensible to the large font size since the corner become sparse. The color based methods assume that text has uniform color. They locate text quickly and are efficient when the background mainly contains uniform regions but they faces difficulties when the text is embedded in complex background or touches other graphical objects. The texture-based methods are used sometimes in compressed domains or combined with machine learning techniques. These methods assume that the text regions have special texture pattern different from other object of background. They usually divide the image into blocks. They employ at first corners [15], High-frequency wavelet coefficients [12], Gabor filter [37] and high-level central moments [20, 42] as texture descriptors of each blocks in order to characterize patterns that belong to textual information. Then a classification into text and non text bloc stage is performed using k-means clustering [12], neural network [37] and SVM [20]. These methods are insensitive to background colors but they find difficulties when the background contains textures that have similar periodic structures as the text region. In addition, it is hard to find accurate boundaries especially when dealing with small text size. Li et al. [21] use mean, second- and third-order central moments in wavelet domain as the texture features and a neural network classifier is applied for text block detection. Sin et al. [34] detect the text using features calculated on the autocorrelation function of a scanline. They assume that text is long and enclosed by a rectangular frame as in the panel of the billboard. Furthermore, several scanlines of a text rectangle are concatenated, which creates problems due to the non-alignment of the phases of the Fourier-transforms of the different scanlines. The method cannot be applied to short text. Wu Manmatha [42] combine the search for vertical edges with a texture filter to detect text. These binary edge clustering techniques are sensitive to the binarization step of the edge detectors. Qian et al. [30] use a DCT coefficient. Candidate text blocks are detected in terms of block texture constraint. An adaptive method for the horizontal and vertical aligned text lines determination is designed. The block regions are further verified by local block texture constraints. The text block region is localized by the horizontal and vertical block texture projections. Y.Zhong et al. [47] extract text directly from JPEG images or MPEG video. Texture characteristics computed from DCT coefficients are used to identify $8 \times 8$ DCT blocks that contain text. For each DCT block, it computes the energy of horizontal text. Morphological operations are applied to remove the isolated noise blocks and merge disconnected text blocks. Liu et al uses [23] firstly the variances and covariancs on the wavelet coefficients of different color channels as color textural features to characterize text and non-text areas. Secondly, the k-means algorithm is chosen to classify the image into text candidates and background. Finally, the detected text candidates are verified by a set of rules and localized by the projection profile. Tekinalp et al. [37] propose two main steps text detection, the coarse and the fine detection. Within the first step a multiscale wavelet energy feature is employed to locate all possible text pixels and then a density-based region growing method is developed to connect these pixels into text lines. In the second step four kinds of texture features are combined to represent a text line and a SVM classifier is employed to identify texts from the candidate ones. Kim et al [19] used a support vector machine(SVM) to analyze the textural properties of texts. the intensities of the raw pixels that characterize the textural pattern are passed directly to the SVM. Text regions are identified by applying a continuously adaptive mean shift algorithm (CAMSHIFT) to the results of the texture analysis. This method is more effective and faster than general texture-based methods but it may have scale problem due to CAMSHIFT. The texture based approaches are generally time consuming especially those in compressed domain. They obtain good results for large text.Further more it's difficult to find a discriminative feature which can distinguish text from non-text-like texture pattern. The compressed domain methods i.e. wavelet based methods MPEG 2, MPEG 4 may be quickly obsolete due to the rapid evolution in the field of video coding domain. The edge-based methods detect text region under the assumption of the density in stroke and contrast characteristics of text. This kind of approaches can hardly handle large-size text or with complex background. Ngo et al [26] proposed a detection technique that is based on background complexities analysis. Video frames are classified into four types according to the edge densities. Effective operators such as the repeated shifting and smoothing operations are applied for the noise removal of images with high edge density. This method depends on the global threshold set to suppress the non edge pixels which may remove possible pixels text regions. Wong et al in [41] compute maximum gradient difference (MGD) to detect potential text line segments from horizontal scan lines of the video frames. Potential text line segments are then expanded or combined with potential text line segments from adjacent scan lines to form text blocks, which are then subject to filtering and refinement. This method is sensible to objects similar to text and which present also high contrast with the background. Wolf et al [40] used sobel operator, a measure of accumulated gradients within a window and morphological post processing to detect the static text. Chen et al [7] apply canny filter, then according to the direction of edge they apply morphologic operators in horizontal and vertical direction to get text blocks characterized by short horizontal and vertical edges connected to each other. Obtained region are then verified by the SVM classifier with CGV feature. Lyu et al.[24]extracted the edge maps by use of four directional sobel edge detection operators, a local thresholding technique, hysteresis edge recovery and a coarse-to-fine localization scheme to identify text regions based on multi-level detection. The final text detection results are the integration of those detections from different levels. Sato et al[33] proposed horizontal differential filter to extract edge followed by setting constraints as size, fill factor and horizontal-vertical aspect ratio to detect text region. Tsai et al [38] method assumes that the scrolling text only exists in the boundary of the image and considers only Chinese text. They apply Sobel edge detector in two directions. Then a threshold is determined by computing the mean vertical/horizontal intensity of the entire edge frame. Geometric constraints and projection profile are applied to locate the Chinese scrolling text. Xu et al [43] used edge density feature within a sliding window of different size. If the number of edge pixels in a window is larger than a predefined threshold it is considered to belong to candidate region. A morphological close operation is applied before labeling the connected component that is grouped together according to a set of geometrical constraints. Final text bounding box are obtained by a refinement module. The method in [46] generates initial connected components by edge detection and morphological operations. Jung et al [18] proposed a text candidate detection based on edge-CCA (connected component analysis), They fuse the N-gray (normalized gray intensity) and CGV (constant gradient variance) SVM classifiers for text candidate verification. Unlike other approach [44, 21, 7] that use SVM to verify if a region is text or not, they use the output score, color distribution and geometric knowledge for refining the initial localized text lines and selecting the best localization result from the different pyramid levels. Based on the edge map Lienhart et al [22] [19] adopted a pre-trained feed-forward network to detect texts in images and video frames. The coarse text regions are detected by a special region-growing method from the edge map images. The coarsely detected regions are refined by horizontal and vertical projections. Nevertheless most of the approaches that have been proposed, extract texts from individual frames independently; while a few approaches exploit the temporal redundancy by using multiple frames integration [22, 14, 22, 32, 40]. More specifically, frame averaging method [14] was introduced to reduce the influence of the complex background. With these methods, an enhanced image for segmentation is ob tained from the average-minimum-maximum pixel value that occurs in each location during the frames containing the same text. Lienhart et al [22] computed the maximum or minimum value at each pixel position over frames. The values of the background pixels that are assumed to have more variance through video sequence will be pushed to black or white while the values of the text pixels are kept. However, this method can only be applied on black or white characters. Wolf et al. [40] assumed that the videotext is static. After the text region is detected, it is localized in the same place in successive frames to reduce the consumption of localization process. Up to our best knowledge, only Tang et al In [36] used temporal information at the detection and localization stage of text within video. They propose a text appearance detection method based on shot boundary detection and feature extraction from frame differences. Four edge filters are applied on the frame difference which is divided into predefined number and size blocks. A vector of features related to each pixels along a fixed number of frames is constructed and passed to a neural network classifier for discriminate text pixels from background. However, this method is sensitive to shot boundary detection.

## III.  A SPATIOTEMPORAL APPROACH FOR VIDEO TEXT EXTRACTION

We propose in this paper a spatiotemporal approach for uncompressed video text localization and identification. The approach starts by detecting potential text regions in a localization step and then validates the effective ones by a filtering process. The edge map of each frame is computed then binarized according to the technique described in[5]. Note that, consecutive video frames are generally similar. An important difference between these two frames is often due to the appearance of new objects that could be eventually text regions. Under this assumption, we calculate difference of each pair of consecutives video frames on which we apply a split-and-merge process. The split step consists of subdividing recursively an edge frame in blocs and measuring its edge density as well as that of its homologous blocs in the next frame. The subdivision is stopped if the edge density of a bloc is less than a predefined threshold or if the size of a bloc is too small i.e. smaller than the size of a readable text. The merge step consists of merging adjacent blocs with similar edge density. The split-and-merge process aims at localizing the new objects (text objects or non text object) which appear within a video frame. Hence, a filtering process is required in order to keep only the effective text regions. Based on the main characteristic as outlined previously we use two criteria in the filtering process: the first measures locally the contrast between a foreground object and the background, the second computes the regularity of line segments forming text regions.

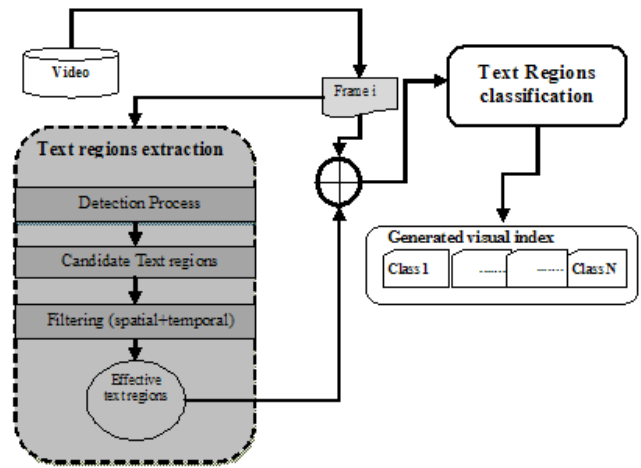

**Fig. 1.** The proposed approach

Text regions issued from the filtering process are identified based on a visual grammar descriptor. The grammar descriptor contains a set of text class features and will be exploited thereafter by our video content-based indexing and browsing system once extracted text regions are classified and identified. With observing of the suggested approach, as shown in figure 1, we note that our approach provides not only video text detection and extraction but also a semantic classification of the already extracted text region.

### A. Potential text region localization

Since the contrast of text is more reliable than the color uniformity, the text detection process mainly depends on the edge information. The Sobel detector [35] is used to detect edges for two reasons: 1)the strokes of all directions are equally treated and 2) it generates double edges, which make text areas much denser than non text areas. For the detection and localization of potential text regions, we compute for each adjacent frame pair the edge map as the square root of the horizontal and vertical Sobel edge map as in equations 1. The obtained edge frames pair is binarized according to the optimal thresolding technique [5] which minimizes the interclass variance of the threshold black and white pixels.

This step aims at getting a binary image that contains only most contrasted pixels.

$$ef_k = \left\| \nabla f(i,j) \right\|_k = \sqrt{f_x^2 + f_y^2} \qquad (1)$$

After the binarization step, we calculate the resulted frame difference between each pair of frames indexed (i, i+1) (see figure 2). Then, a split-and-merge process is applied on each binaries edge frame pair according to the quadtree technique [8] as shown in figure 3 The use of the frame edge difference allows to overcome problems inherent to the background complexity and to optimize extra processing encountered when used a fixed size and number of blocs. For each bloc, we measure the edge density as in equation 2.

$$D = \frac{1}{m \times n} \sum_{x=1}^{m} \sum_{y=1}^{n} I_{edge}(x,y) \qquad (2)$$

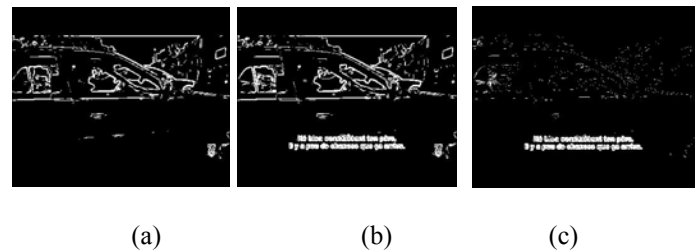

(a)                      (b)                      (c)

**Fig. 1.** Edge frame pair difference (a) frame i, (b) Frame (i+1), (c) frame difference: (b)-(a)

where $I_{edge}(x, y)$ is the intensity value of an edge pixel at the position $x$, $y$ and $m$, $n$ denote respectively the width and the height of the current bloc. Each bloc is divided into sub blocs if its edge density $D$ is greater than a predefined threshold T. The subdivision is stopped if $D$ is less than T or the bloc is too small. In both cases the bloc becomes terminal. Two benefits are issued from the use of density measure: the first is to get a dynamic number of split depending on the edge density value. Blocs having small edge density value (not contains new objects) are not splitted. Hence the number of bloc may vary from a frame to another. The other benefit is the localization of the region of interest. Indeed, the bloc belonging to desired objects presents a high density, especially if it is a text, assuming that the colors of pixels belonging to text strokes are similar. The result of this step is a set of blocs having similar pixel densities at several levels of split. To get the potential text regions the terminal blocs must be grouped within a merge process. The merge process operates by combining the terminal blocs according to their pixels density and their positions. The adjacent blocs with similar pixel densities are merged to form greater blocs and consequently probable text regions.

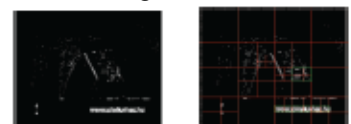

**Fig. 1.** The Split and merge processes by quadtree technique applied on binarized edge frame pair difference

The obtained merged blocs are mapped into the original frame $i + 1$ associated to the current frame pair to extract the real regions of interest which should be filtered within a filtering process.

### B. Text region Filtering

The filtering process is an important task in our approach. The obtained results from the candidate text regions localization process constitute a first localization of the most likely text regions. To reduce false detections of the effective text egions, the filtering process aims at validating each already extracted region and decides whether it is effective or not. This filtering process is based on the size of the detected region, the contrast variation and the regularity of strokes. In the initial filtering step, candidate text region are ignored if they are too small. The remaining ones are then processed in the way described by optimum thresholding method [5]. An analysis of the color spatial variation of all candidate text regions allows us to identify effective text regions. This analysis is based on basic text characteristic which denotes that text characters are generally contrasted with background since artificial text is designed to be read easily. All we need to do is to locate, for each region, pixels belonging to the background and pixels belonging eventually to the text object. Then we determine on the histogram of each thresolded region the positions of the two more dominant peaks $P1$ and $P2$. If the distance $D(P1, P2)$ (see figure 4) between the two peaks is greater than a predefined threshold $\sigma$, the candidate text region is classified as an effective text region, otherwise it will be ignored. I n our experimentation $\sigma = 110$.

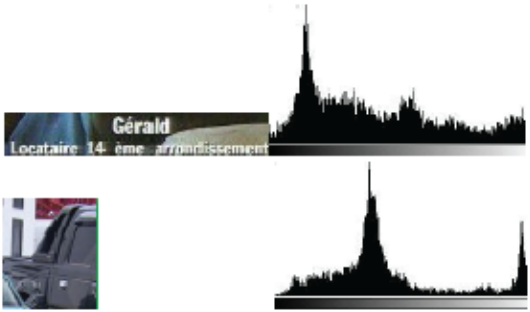

**Fig.4.** Contrast variation of a text region

This filtering criterion fails in the case of regions with a high contrast variation and containing objects other than text. For that, obtained results from this step are filtered again according to a line segment texture feature. The feature is based on the fact that color information is not stable when the illumination changes, so only intensity component will be exploited. Thus, in the gray-scale region image, straight edge content can be approximated by line segments. By using these segments, we proposed a descriptor based a fast Hough Transform principle [6] which represents a perceptual characterization of texture, similar to a human characterization, in terms of regularity, similarity, directionality and alignment as illustrated in figure 5. We remind here that regularity and directionality features are specified in MPEG-7 and more specifically in Texture Browsing Descriptor [28].

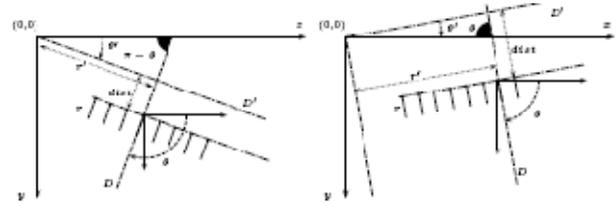

**Fig. 5.** Three-dimensional representation of segments ($\Delta_r$, $\theta$, $r$ and $dist$)

The syntax of our Line Segment Texture Feature, represented in figure 5, is defined as follows:

$$LSTF. \ [\Delta_r, \theta, r, dist] \qquad (3)$$

where each feature represents a characteristic of line region:

- Regularity is represented by $\Delta r$; vector of distances between two adjacent line segments within the same orientation $\theta$.

- Similarity is represented by $r$; vector of lengths of line segments.

- Directionality is represented by $\theta$, vector of angles between the horizontal axis and the perpendicular line of line segments.

- Alignment is represented by $dist$, vector of distances between extremity of line segments and the perpendicular line /of line segments.

One of the most important aspects of our LSTF is the invariance to rotation and the scale changes. For the rotation, we can consider the case in which the angle $\theta$ of line segments has been changed by a variation $\Delta_\theta$, the vector of features becomes:

$$LSTF' = [\Delta_r, \theta', r, dist] = [\Delta_r, \Delta_\Theta + \Theta, r, dist] \qquad (4)$$

On the other hand, the scale change is taking into account by considering a factor $\alpha$. Thus, the feature vector becomes:

$$LSTF'' = [\Delta'_r, \theta, r', dist'] = [\alpha.\Delta_r, \Theta, \alpha.r, \alpha.dist] \qquad (5)$$

In our work, we used the LSTF as a filtering criterion in order to confirm the effectiveness of detected text regions (see figure 6).

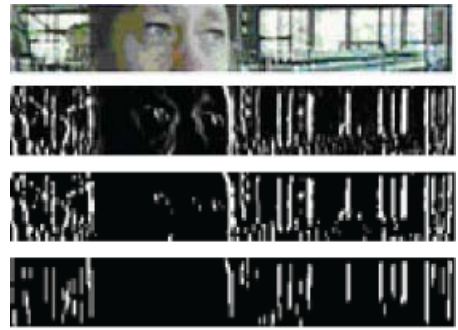

(6.a) Region skipped from CF but filtered by the LSTF

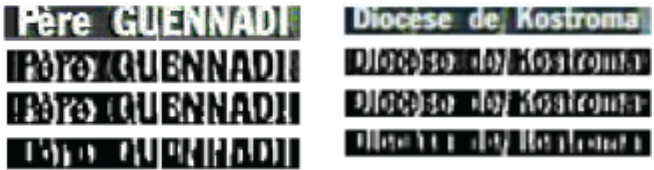

(6.b) Regions kept from CF+LSTF

**Fig.6.** The Extraction of LSTF of potential text region after filtering by CF Hence, in a given orientation $\theta$ (ie. $\theta = 90\circ$ for horizontal text) and a similar text size (similar length $r$ of segments forming line text region) we compute the regularity normalized component as in equation 6, 7.

$$R = \frac{\sum_{i = first}^{last-1} \left| (l_{i+1} - l_i) - \overline{d} \right|}{(ns-1)*d}.$$  **(6)**

Where

$$\overline{d} = \frac{last - 1 - first}{(ns-1)}.$$  **(7)**

with $last$ and $first$ represent respectively the position of the last and the first detected line segment within text region. $ns$ is the total number of line segment within text region. The $R$ metric returns a value in the range [0, 1]. The higher the value, the higher probability to get a text region is. A region is considered as text if the value of $R$ is less or equals than a fixed threshold, otherwise this region is ignored (in our experiments $\overline{R} = 0.2$).

## IV. AUTOMATIC TEXT REGION IDENTIFICATION FOR VIDEO BROWSING

The role of text detection and localization is to find the image regions containing only text that can be directly highlighted to the user or passed to an **O**ptical **C**haracter **R**ecognition system (**OCR**) in order to get the plain text. Indeed, localized text regions are meaningful by themselves. For example, finding the appearance of a caption in news video can help to locate the beginning of a news event and enable video browsing by the textual image. Once text regions are recognized by OCR module, applications such as keyword-based video search become possible. Unfortunately, finding an efficient video text recognition system is still a challenging problem because, on one hand, text images have usually complex background, noise, illumination and variation. On the other hand existing OCR techniques are designed for images where the color of text is clearly different from that of background and especially for scanned document images presenting a simple color distribution and high resolution (200dpi-300dpi). Moreover, the success of these approaches is limited by the language supported by OCRs, so the extracted text is useless if there are no OCRs in the considered language. Consequently, theses drawbacks may prevent these techniques from practical applications like video browsing by content. To overcome these problems, we present an independent language technique for text region classification based on video-grammar.

### A. Grammar initialization

In TV brodcasting, each TV channels define its own style that allows the creation of its visual identity or that of a program. This style is defined by audiovisual elements reflecting the identity of a video sequences. These elements are in general recurring; their application on the video products follows often rules defined as a "graphic charter" and represented in our work by a visual grammar. For exemple the news topic text of M6 tv channel are always framed in the bottom by a black box. In this section, we present a video text image classification approach based on specific visual elements (forms, colors) that appear with text. Thus, to identify and classify the various kind of text within these programs, we use a visual grammar which characterizes all instances of the same TV program. This grammar allows representing the visual recurrent elements for each text classes using a set of descriptors appropriate for each TV channel and each program. This grammar can be expressed as in equation 8.

$$G = <C, F_{CTi} > .$$  **(8)**

To characterize each class $CTi$ of video text regions, it is important to use a discriminate feature $F_{CTi}$ of each class of these regions.

To characterize each class $CTi$ of video text regions, it is important to use a discriminate feature $FCTi$ of each class of these regions. In this work, we choose a colorimetric descriptor to characterize the visual element of the text regions classes. Indeed, the dominant color [31, 29] is one of the most relevant descriptors to represent visual information and characterize a homogeneous image area. So, all the occurrences of the same text-classes present a similar color descriptor in a video sequence as shown in figures 7, 8. For example, the texts representing the names of cautioned soccer players are generally accompanied by a homogeneous yellow area color at the same position to indicate the yellow card. The visual grammar descriptor associated to a video sequence contains a set of descriptive and discriminated features of text classes. The generation of this grammar is done in a semi-automatic way. Indeed the spatial localization of the image region representing the visual grammar is done manually by a simple use of a mouse. While the extraction and the generations of colorimetric descriptor are done automatically.

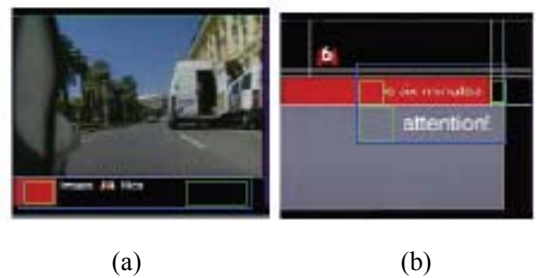

(a)                    (b)

**Fig.7.** Sample of visual invariants of ''M6'' le six minutes program. (a) Text region descriptors of comments (b) Text region descriptors of subject titles

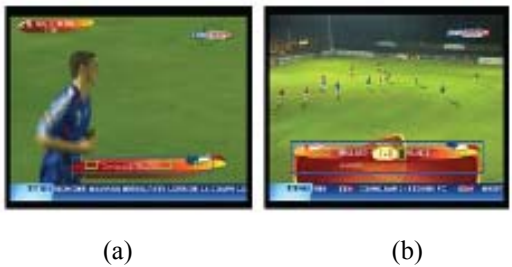

**Fig.8.** Sample of visual invariants of "eurosport" TV channel soccer play (a) Text region descriptorsof player names (b) Text region descriptors of the score.

Hence, a text class $C_{Ti}$ is described by a set of features, expressed by the equation 9, which include firstly the position and the size of the region (**blue** rectangles) in which text will appear and secondly, a set of discriminated sub regions specific to each text class $\vec{v}_i$ (drown by **green** rectangles) as in figure 7, 8.

$$F_{CTi} = \left\{ X_i, Y_i, W_i, H_i, \vec{V}_i \right\}. \qquad (9)$$

where $Xi$ and $Yi$ represent the coordinates of the first diagonal. $Wi$ and $Hi$ represent its width and height respectively. The $SRi$ is a vector of $k$ sub-regions characteristics of the text class $CTi$ as indicated by the equation 10.

$$\vec{V}_i = \left[ SR_{ij} \right] = \left[ x_{ij}, y_{ij}, w_{ij}, h_{ij}, H_{ij}, S_{ij}, V_{ij} \right]. \qquad (10)$$

Where $x_{ij}$ and $y_{ij}$ represents the coordinates of the first diagonal of the sub-region $j$ associated to the text class $i$, $w_{ij}$, $h_{ij}$ represent its width and height respectively. $H_{ij}$, $S_{ij}$, $V_{ij}$ represent respectively the hue, the saturation and the value of the dominant color of the sub-region number $j$ computed according to the equation 11 based on normalized $R$, $G$, $B$ values ($R, G, B \in [0, 1]$) of each pixel belonging to a sub-region. In fact, the RGB color space is non uniform and consequently two visually similar colors may lead to a significant Euclidian distance when computing the difference between them. So, to overcome this problem and in order to get a faithful representation of the perceptual color and its associated value we used the HSV model [18].

$$\psi_{ij} = \frac{1}{W_{ij} \times H_{ij}} \sum_{x=1}^{w} \sum_{y=1}^{h} \psi(x,y). \qquad (11)$$

with $W$, $H$ represent respectively the width and the height of the sub-region $j$ of a text class $i$; $\Psi = H$, $S$, $V$ represents respectively the hue, the saturation and the value of the sub-region pixel at the position $(x, y)$.

Hence, the Grammar Specifier (GS) selects and delimits the representative region of a text class. Thereafter, he determines the representative sub-regions of the visual invariants for this region represented by the vector of sub-region features $\vec{v}_i$. Geometrical properties (coordinates, the height and the width) and the $H, S$ and $V$ component of the dominant color are extracted and stored within a descriptor file.

## B. Text image classification

The classification process of a text region extracted from video sequences provides a key feature for video content indexing and browsing. So when a text region is extracted and classified it can be used as a high level index for direct content access. The classification process is performed as in the pseudo-code of Algorithm 1. Based on the predefined grammar-descriptors, the similarity between features of each text class $CTi$ and the ones of the query extracted text region $Q$ is measured. First, a spatial mapping between the extracted text regions (figure 9.a.) and the area specified by the **GS (grammar specifier)** as the region in which the text may be incrusted (see figure 9.b) is made. Secondly, the similarity between the GS'specified sub-region features on one hand and those extracted at the same position on the other hand is measured as in equation 12. In the spatial mapping between the automatically localized text region (region B) and the specified user text region (region A) four cases are possible (see figure 10):

a. Equality: the two areas have same dimensions and the same coordinates.

b. Covering: the region B is entirely included in region A.

c. Partial overlapping: the two areas have a common part (area of overlap).

d. Disjunction: there is no overlap between the two areas.

```
Algorithm 1 : text region identification
for each text region already extracted do
    for each text class do
        compute the overlap area between the text region and the text
        class region;
        if ((covering)or(Equality)or(partial overlap)) then
            compute the distance D_i (SR_ij, Q_j);
        else
            perform next text class CT_i;
        end
    end
    if Min(D_k (SR_ij, Q_j)) ≥ α then
        classify the current region in the class i ;
    end
end
```

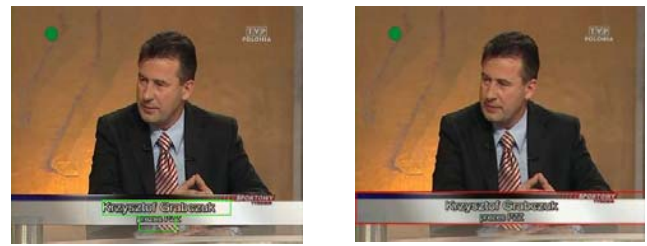

(a) automatic detected text Region        (b) specified GS text region

**Fig.9.** Text regions mapping

For the first two cases (equality, covering) the text region is retained and then identified. In these two cases, we chose to retain the coordinates of region A instead of those of the region B since the later is detected automatically. Whereas the coordinates of region A are defined by the GS who must choose the good boundaries of the regions that he wishes to detect. In the case of partial overlapping, only the regions whose their overlapping area exceeds a definite rate of the total area are kept, the others are considered as non identifiable regions. For the disjunction case of the two regions, the region B contains text which is not identifiable according to text classes defined by the grammar and then this region is ignored. After this partially identified regions, we obtain a set of regions which should be identified definitively based on the sub-regions features $\overline{v}_i$. Consequently, we need to define a metric measuring the distance between each sub-region within $\overline{v}_i$ of the query localized text region $Q$ and cases from the text regions class $CT_k$. One of the most popular choices to measure this distance is known as Euclidean (12).

$$D_i(SR_{ij}, Q_j) = \sqrt{(H_{SRij} - H_{Qj})^2 + (S_{SRij} - S_{Qj})^2 + (V_{SRij} - V_{Qj})^2} \quad (12)$$

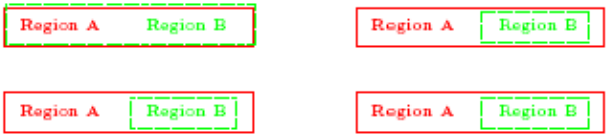

**Fig.10.** Spatial mapping between text regions detected automatically and those specified by GS

where $D_i$ is the distance between the sub-region number $j$ associated to the text region class number $i$ ($SR_{ij}$), already stored within grammar descriptors and its homologous query sub-region associated to the automatically localized text region (Q).Then a mean similarity distance $\overline{D_i} \in \{D_1, D_2, ... D_n\}$ is computed as shown in equation 13 and assigned to each text class $CT_i \in \{CT_1, CT_2, ..., CT_n\}$.

$$\overline{D_i(SR_{ij}, Q_j)} = \frac{\sum\limits_{j=1}^{k} \sqrt{(H_{SRij} - H_{qj})^2 + (S_{SRij} - S_{qj})^2 + (V_{SRij} - V_{qj})^2}}{k} \quad (13)$$

With $k$ the number of the text class sub-regions. Hence, a table of similarity distance $\overline{D_i}$ ($SR_{ij}, Q_j$) is constructed between all possible pairings of query text region ($Q_j$ (automatically detected) and grammar text regions class $CT_i$. The identification process is achieved by choosing the associated grammar to the video sequences. A query text region is considered as an effective text instance of the class $CT_i$ only if it presents the smaller distance $\overline{D_i}$ ($SR_{ij}, Q_j$) already stored in the table of similarity and greater than a factor $\alpha$ verifying $Min(\overline{D_i}$ ($SR_{ij}, Q_j)) \geq \alpha$.

### C. Video TOC generation

Video browsing and information retrieval in multimedia documents are still challenging problems. In fact, efficient access to video semantic units is difficult due to their length and unstructured format. These difficulties motivated several researchers to develop more robust video browsing systems to facilitate user's access to videos semantic units. The basic idea in these systems is to generate a **T**able **O**f **C**ontent (**TOC**), for the video files, composed by anchors to the highlight video segments. Motivated by the above facts, we propose anautomatic construction of the video TOCs as visual indexes for the video documents. After the extraction and identification processes, we gather the resulted text regions in a set of thematic TOCs. The semantic topics corresponding to different TOCs are deduced from grammars that classify semantically the kind of text-regions appearing in the video. This classification allows grouping the text regions in semantic classes (e.g., news titles, sport results). Thus, all text instances of each class constitute a visual index representing its topics. Each text region represents an anchor that points to the semantic unit described by this text.

## V. EXPERIMENTAL RESULTS

In this section, we present the experimental results for the evaluation of our AViTExt system dedicated for video browsing by image text content. Our experiments are conducted on sixteen video sequences having a 352x288 frame size with a frame rate of 25fps. Some of these sequences are captured with Pinnacle PCTV-Sat card, others are downloaded at *http://www.informatik.unimannheim.de/pi4/lib/projects/moca/* and *http://www.youtube.com*.These sequences are portions of news broadcasts, sports program and fiction movies of *drdishTV, m2TV, TVP Polonia, Fox 5, M6, RAI SPORT sat, TpsFoot, Reuters, CNN, BBC and TF1* channels. Within this test database, text appears in a variety of locations, colors, sizes, fonts and background complexity. Since the evaluation criteria for the text detection and localization and those for the text classification are different, we divide the experiments into two phases. The first phase focuses on the text detection and localization algorithms, which take a video clip as input and extract the localized rectangular text regions. The second phase evaluates the text identification algorithm, which classifies the already extracted text region by the first step into semantic class according to the grammar descriptors, and produces a table of content (TOC) containing image text region grouped into semantic class.

### A. Evaluation of Text Detection and Localization

The experimentation process deals with All static regions distinguishable as text by humans. Closely spaced words lying along the same alignment are considered as the same text occurrence. Overall, test data contains 1132 temporally artificial text occurrences (357 news, 392 sports, 383 film). For a quantitative evaluation, we define that a detected text region is correct on the condition that the intersection of the detected text region and the ground-truth text region exceed 85%. The ground-truth text regions in the testing video sequences are localized manually. Thus, we define a set of quantitative evaluation measurements. Recall rate evaluates how many percents of all ground-truth video text regions are correctly detected. The Precision rate evaluates how many percents of the detected video text regions are correct. False alarm rate evaluates how many percents of the detected videotext regions are wrong as defined in (14), (15) and (16).

$$Recall = \frac{Number\ of\ correted\ detected\ text\ regions}{Number\ of\ all\ ground\_truth\ text\ regions} \quad (14)$$

$$Precision = \frac{Number\ of\ correted\ detected\ text\ regions}{Number\ of\ detected\ text\ regions} \quad (15)$$

$$False\ alarm = \frac{Number\ of\ wrongly\ detected\ text\ regions}{Number\ of\ detected\ text\ regions} \quad (16)$$

The efficiency of our approach for video text extraction comes from the use of two main steps : potential text region detection based on a quadtree technique and text region filtering. This latest step is based on two significant features of video text regions: the contrast feature (CF) and the line segment texture feature (LSTF). The overall results of our automatic video text extraction were 93.55% of Recall, 91.53% of precision and 8.47% as rate of false alarm.

TABLE 1: EXPERIMENTAL RESULTS OF THE TEXT REGION EXTRACTION

| Measures | CF | CF+LSTF |
|---|---|---|
| Recall | 93.90% | 93.55% |
| Precision | 75.66% | 91.53% |
| False alarm | 24.34% | 8.47% |

The results presented in table1 show that the use of the two features on the filtering process improve notably the precision of the overall approach rising up from 75.66% by using only the contrast feature to 91.53% when applying the LST feature. In fact, the potential text region localization step detects all new objects appearing between each pair of frames. However, the CF filtering criterion fails in the case of regions having a high contrast variation and contains other objects than text as shown in figure 6. Nevertheless, the LSTF decrease the false detection. These results as illustrated in figure11 show that our technique is efficient, capable to locate text regions with different character sizes and orientation, even in case of texts with law contrast or occurring within complex image background and that lay on the boundary of the video frame. Besides, our algorithm has the high-speed advantage with two frame pair per second in the frame size of 352 x 288 which suitable for videotext extraction and better than the speed of the algorithm in [24] which according to our implementation perform 1 frame per second.

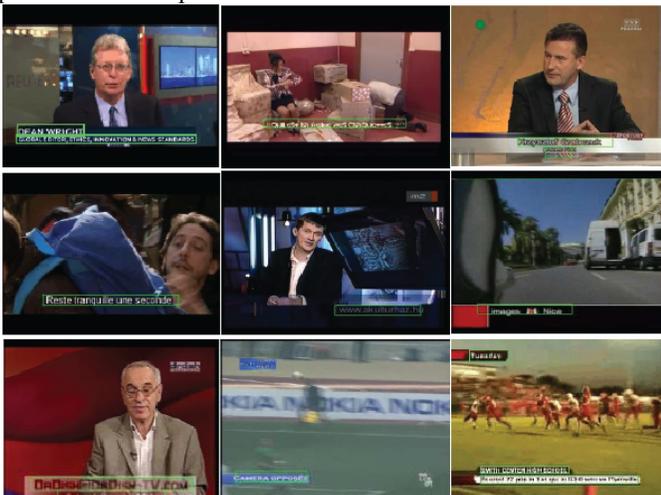

**Fig.11.** Text region Location Examples

Furthermore, the extracted text regions are precise when text strokes occupy extended area. The missing detections are mainly caused by the following two cases: (1) the contrast between text and background is very low and (2) the gradual appearance of text. The false detections are due to the fact that texture information of background regions similar to that of text regions. We have carried out comparison with other interesting methods presented respectively by Liu et al [24] and lienhart et al [22] as depicted in table 2.

TABLE 2: COMPARISON FOR PRECISION, RECALL AND FALSE ALARM

| Measures | Liu [24] | Lienhart[2] | Our method |
|---|---|---|---|
| Recall | 68,40% | 91.02% | 93.55% |
| Precision | 84,44% | 86.25% | 91.53% |
| False alarm | 21,66% | 15.47% | 8.47% |

However, we note that the comparison with these two methods is very difficult due to the lack of a common video test database and that the referred methods consider the video as a set of independent frames. We note then that direct omparison shows that the use of our LSTF features increase notably the precision of our approach compared with those in [24] and [22] and decrease consequently the rate of false alarm.

*B. Evaluation of Text Region classification process*

For the evaluation of the classification process, we defined two metrics described by (17) and (18) :

$$TxTI = \frac{NbTI}{TRTo} \quad (17)$$

$$TxTNI = \frac{TRTo - NbTI}{TRTo} \quad (18)$$

with :

*TxTI*: rate of text regions detected and identified.
*NbTI*: number of text regions detected and identified.
*TxTNI*:rate of text regions detected and not identified.
*TRTo*: total number of automatically extracted text regions.

TABLE 3: TEXT REGION CLASSIFICATION

| Metrics | VALUE |
|---|---|
| *TxTI* | 92,59% |
| *TxTNI* | 7,41% |

From Table 3, we can see that our method achieves significantly high rate *TxTI* of identified text regions rising 92.59% of text regions. Some already extratcted text region may be misidentified due to the illusion between subregions characterizing the text region specified in the grammar descriptor and the homologous in the automatically extracted region.

## VI. CONCLUSION AND FUTURE WORK

In this paper, we proposed an approach for video text detection, localization and identification. This approach constitutes a contribution to the video content-based indexing field and operates in two mainly steps: text region extraction and text region identification.With the text region extraction process, we localize the candidate text regions, according a split and merge processes applied on each binarized edge frames pair within video sequences. The resulting extracted regions are then filtered, according to their size, contrast variation and regularity in order to get the effective ones. The identification process aims at classifying the effective text regions in semantic classes and publishes them within visual indexes providing easy browse of video content. First, we define the various kinds of text to be extracted and their visual characteristics to form the signatures which will be stored within a grammar descriptor as text region class features. Then, we identify the already extracted text regions from the video sequences by comparing their signatures with those already stored in the grammars descriptor. This step allow the generation of the index containing the image text regions enabling consequently video content browsing. Many applications can be derived from this automatic text locating technique. For instance, automatic video summarization and automatic generation of video content table enabling further video browsing and retrieval. In this context, we developed a prototype called "AViTExt: Automatic Video Text Extraction". This prototype (See figure 12) can be used to browse video sequences by text images regions extracted automatically according to our proposed approach. It includes three principals' components. The first component named "Tree views" provide a hierarchical view of video sequence as a table of content (TOCs) where each node present a class of text region. The second one named "Visulazor" is a screen used to visualize scenes within all localized text index. The third component is a keyword based search engine which will be explored in our future work when we convert all detected text region to its associated ASCII text by the use of OCRs techniques. Currently, we only provide a text detection localization and classification method. Text should be clearly extracted from its background to obtain a good recognition result for the characters. Special technique should be investigated to segment the characters from their background as in[48] before putting them to an OCR module. Motivated by the efficiency and the novelty of our approach for extraction of still text, which gives the most pertinent information about the content, we focus our future work on the determination of splitting and merging criteria that should be adaptive in step with the complexity of the background of the video frames. A known limitation of our approach is that it cannot detect rightly moving texts due to the assumption of stationary text. However, this capability can be enabled by tracking the detected text region within a temporal window.

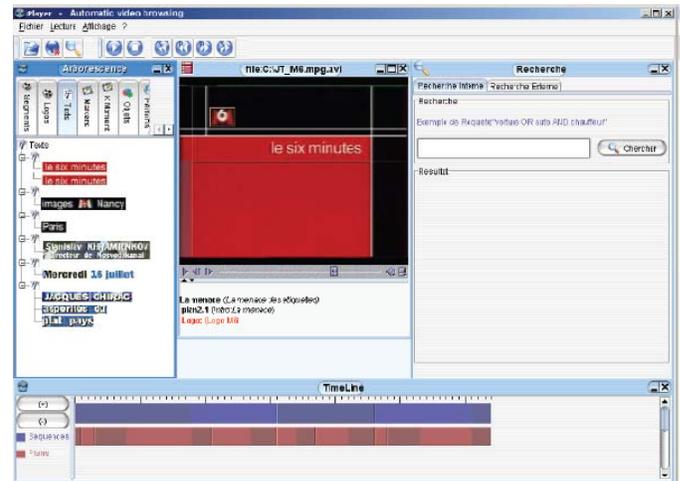

**Fig.12.** Main Interface of AViTExt

Our investigation will also concern the integration of the video grammar as an interesting way to guide the recognition of text regions. Consequently, our prototype will enable, in addition to our current video content browsing by text images, a full text search.

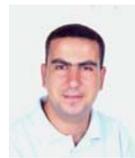
**Bassem BOUAZIZ** is preparing a Ph.d degree in Computer Science at the *Sfax University*, Tunisia. he is a Teaching Assistant at the Higher Institute of Computer Science and Multimedia, Sfax, Tunisia. His researches concern image and video processing, especially the video content structuring and indexing and browsing.

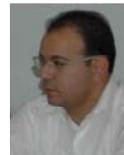
**Walid MAHDI** received a Ph.d. in Computer and Information Science from *Ecole Centrale de Lyon, France* in 2001. He is currently Assistant Professor at Higher Institute of Computer Science and Multimedia, at the University of Sfax, Tunisia. His research is about image and video processing.

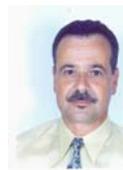
**Abdelmajid Ben-Hamadou** is Professor in Computer Science and the director of the multimedia and advanced computing laboratory in the higher institute of computer science and Multimedia, Sfax, Tunisia. He has been the director of the LARIS reserach unit since 1983. His research interests include automatic processing of Natural Language, object-oriented design and component software specification and image and video processing.